# Search for New Gauge Boson $Z'_{B-L}$ at LHC using Monte Carlo simulation

**H.M.M.Mansour[1,2], Nady Bakhet[1,3]**

(Affiliation): [1]Department of Physics, Faculty of Science, Cairo University
Email: [2]mansourhesham@yahoo.com, [3]nady.bakhet@cern.ch, nady.bakhet@ yahoo.com

**Abstract**

In the present work we search for $Z'_{B-L}$ heavy neutral massive boson in the dielectron events produced in proton-proton collisions at LHC. To detect $Z'_{B-L}$ at LHC we used the data which is produced from pp collisions by Monte Carlo events generator for different energies at LHC then we use the angular distribution, invariant mass, combined transverse momentum and combined rapidity distributions for the dielectron produced from $Z'_{B-L}$ decay channel to detect the $Z'_{B-L}$ signal. B-L extension of the SM model predicts the existence of a $Z'_{B-L}$ heavy neutral massive boson at high energies. From our results which we had simulated for $Z'_{B-L}$ in the B-L extension of standard model, we predict a possible existence of new gauge $Z'_{B-L}$ at LHC in the mass range 1 TeV to 1.5 TeV.

**Keywords:**

## I. INTRODUCTION

The fact that the neutrinos are massive indicates that the Standard Model (SM) requires an extension. B-L model is an extension for the SM which is based on the gauge group [1-3] $G_{B-L} = SU(3)_C \times SU(2)_L \times U(1)_Y \times U(1)_{B-L}$. The invariance of the Lagrangian under this gauge symmetry implies the existence of a new gauge boson (beyond the SM ones) and the spontaneous symmetry breaking in this model provides a natural explanation for the presence of three right-handed neutrinos in addition to an extra gauge boson and a new scalar Higgs. Therefore, it can lead to a very interesting phenomenology which is different from the SM results and it can be tested at the LHC.

An extra neutral massive gauge boson corresponding to the B-L gauge symmetry is predicted. There are many models which contain extra gauge bosons. These models can be classified into two categories depending on whether or not they arise in a GUT scenario. In some of these models, Z' and SM Z are not true mass due to mixing. This mixing induces the couplings between the extra Z' boson and the SM fermions.

In the B-L extension of the SM model, the extra $Z'_{B-L}$ boson and SM fermions are coupled through the non-vanishing B-L quantum numbers. Searching for $Z'_{B-L}$ is accessible via a clean dilepton signal at LHC. We will simulate B-L extension of the SM at LHC which is based on the gauge group $G_{B-L} = SU(3)_C \times SU(2)_L \times U(1)_Y \times U(1)_{B-L}$ using MC programs then search for the $Z'_{B-L}$ boson in the dielectron events produced in pp collisions at different energies of LHC where the leptonic decay $Z'_{B-L} \to l^+ + l^-$ provides the most distinctive signature for observing the $Z'_{B-L}$ signal at the Large Hadron Collider. The results in this paper were produced by using simulation events generator PYTHIA8 [4-6] and other software tools as CalcHep, MadGraph/Madevent, FeynRules, ROOT data analysis, Physics Analysis Workstation (PAW), ROOFIT package to fit any resulted histogram in order to get P.D.F. (Probability density function) and Mathematica.

In this paper the results are organized into three subsections. First, we present the production of $Z'_{B-L}$ at LHC which includes production cross section, different branching ratios and total width. Secondly, the detection of $Z'_{B-L}$ signal at LHC via the decay channel $Z'_{B-L} \to e^+ + e^-$ then we study the dielectron angular distribution, dielectron asymmetry, Drell Yan background events for this channel and dielectron invariant mass. Thirdly, we present the properties of $Z'_{B-L}$ which include Luminosity, Significance, Transverse momentum and Rapidity.

## II. RESULTS

In this section we present our results for production and detection of a $Z'_{B-L}$ signal of the B-L extension of SM using MC programs [7-8]. We present the production cross section of $Z'_{B-L}$ at LHC as a function of $Z'_{B-L}$ mass for various g" values (where g" is the U(1)$_{B-L}$ gauge coupling constant) and for various energies at LHC, branching ratios as a function of $Z'_{B-L}$ mass for heavy neutrino mass = 200 GeV which will affect the results of $Z'_{B-L}$ due to the fact that it is a heavy particle. We obtained different results in comparison with Ref [9]. The analysis in this paper did not take into account the existence of new decay channel for heavy neutrino in B-L model which is one of the important signatures of B-L model and they did not give any branching ratio for heavy neutrino. After that we will present $Z'_{B-L}$ total width as a function of $Z'_{B-L}$ mass for various values of g". Both angular distribution of dielectron and invariant mass of dielectron produced of $Z'_{B-L}$ decay are used to detect $Z'_{B-L}$ signal at LHC. Finally, we will focus on the properties of $Z'_{B-L}$ such as Luminosity, Significance, Transverse momentum and Rapidity.

### A. Production of $Z'_{B-L}$ at LHC

*1. Production cross section*

In figures 1 and 2 we present the production cross section for $Z'_{B-L}$ for the most relevant production mechanisms for different CM energies. Figure 1 gives the cross section for $Z'_{B-L}$ at LHC as a function of $Z'_{B-L}$ mass for various g" values (where g" is the U(1)B.L gauge coupling constant) at CM energy of LHC = 14 TeV. Figure 2 gives cross sections for $Z'_{B-L}$ at LHC for CM energies √s = 5, 7, 10, 12, 14 TeV at fixed value of g" = 0.2. At the patron level, the $Z'_{B-L}$ production cross section depends on two main parameters, the mass of $Z'_{B-L}$ and the coupling constant g". Therefore, the B-L model is controlled by two parameters: the mass of the $Z'_{B-L}$ and the coupling constant g" determining $Z'_{B-L}$ couplings.

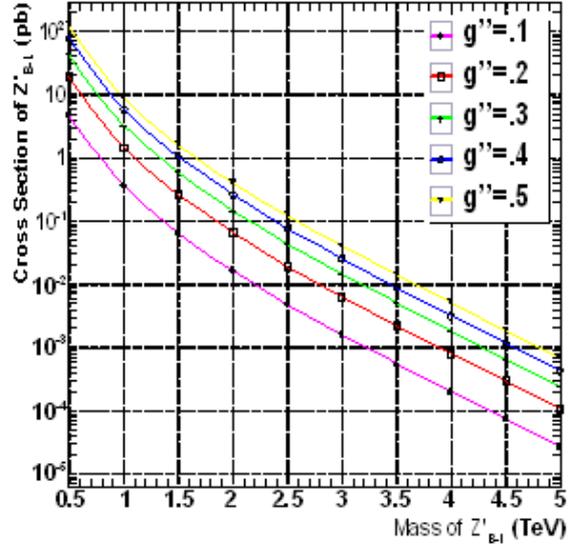

**FIG. 1.** Cross section for $Z'_{B-L}$ as a function of $Z'_{B-L}$ mass for various g" values (where g" is the U(1)$_{B-L}$ gauge coupling constant) at fixed CM energy of LHC=14 TeV

Two experimental constraints exist on these two parameters, the first comes from direct search for heavy neutral gauge bosons at the CDF experiment which excludes a $Z'_{B-L}$ mass less than 600 GeV and the second limit comes from LEP II where:

$$\frac{Z'_{B-L}}{g''} > 6 \text{ TeV} \qquad (1)$$

The interactions of $Z'_{B-L}$ boson with the SM fermions in B-L model is described by

$$L^{Z'_{B-L}}_{int} = \sum_f Y^f_{B-L} g'' Z'_\mu \bar{f}\gamma^\mu f \qquad (2)$$

Where $Y_{B-L}$ is the B-L charge associated with the fermions f (see table 1). The extra neutral gauge boson $Z'_{B-L}$ acquires a mass due to the B-L gauge symmetry breaking

$$M^2_{Z'_{B-L}} = 4g''^2 v'^2 \qquad (3)$$

where g" is the U(1)$_{B-L}$ gauge coupling constant and $v'$ is the symmetry breaking scale

| Particle | $Y_{B-L}$ |
|---|---|
| $L$ | $-1$ |
| $e_R$ | $-1$ |
| $v_R$ | $-1$ |
| $q$ | $1/3$ |

**TABLE I.** B-L quantum number for different particles.

The production cross sections for $Z'_{B-L}$ signal in figures 1 and 2 were computed using MadGraph5 and PYTHIA8 where we generated the process pp → $Z_{B-L}$ of B-L model using MadGraph5 and export this process to PHYTHIA8 then the main switches are on for Initial state Radiation(ISR), Final State Radiation(FSR), FSRinResonances, Decay hadronization, allow resonance decays and master switch for multiparton interactions to stay on.

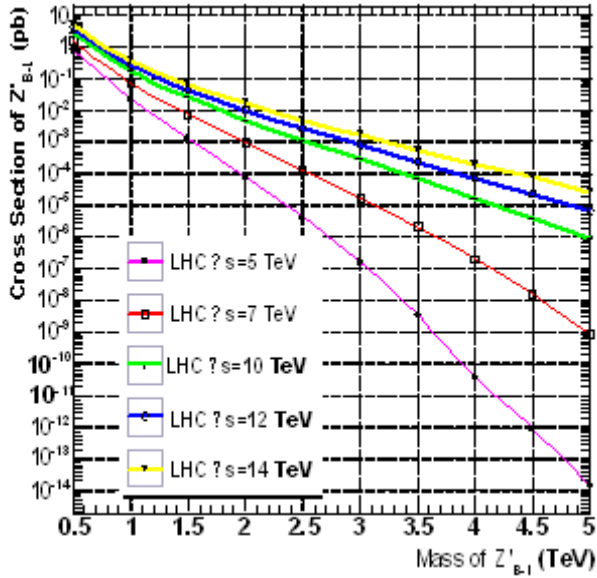

**FIG. 2.** Cross section for $Z'_{B-L}$ as a function of $Z'_{B-L}$ mass for various energies of LHC at √s = 5, 7, 10, 12, 14 TeV at fixed value of g" =0.2

### 2. Branching ratios of $Z'_{B-L}$

In figure 3, the branching ratios of $Z'_{B-L}$ to different quarks are equal approximately and for different leptons are higher than those for quarks. Also the branching ratio for heavy neutrino which have mass 200 GeV in our search is less than those for the charged leptons and light neutrinos. In particular, BR($Z'_{B-L} \to l^- l^+$) varies between 17% to 17.5% where (l = electron, muon, tau) but for heavy neutrino BR($Z'_{B-L} \to h_v h_v$) varies from 6% to 8% and for light neutrino BR($Z'_{B-L} \to v v$) varies between 8.5% to 9% and BR($Z'_{B-L} \to q q$) varies from 5.5% to 6%. The probabilities that $Z'_{B-L}$ can decay into one light and one heavy neutrino are highly suppressed by the corresponding (heavy-light) neutrino mixing and thus they can safely be neglected. Heavy neutrino is the most characteristic for B-L model so it has an effect on different branching ratios because it is rather massive than the SM neutrino. From figure 3 one can search for $Z'_{B-L}$ at LHC via a clean dilepton signal which can be one of the first new physics signatures to be observed at the LHC. We will study $Z'_{B-L}$ in this paper by using the decay channel of $Z'_{B-L}$ to electrons pair using PYTHIA8 and turn off all other decay channels for $Z'_{B-L}$ where the ratio of dielctron channel is the highest one.

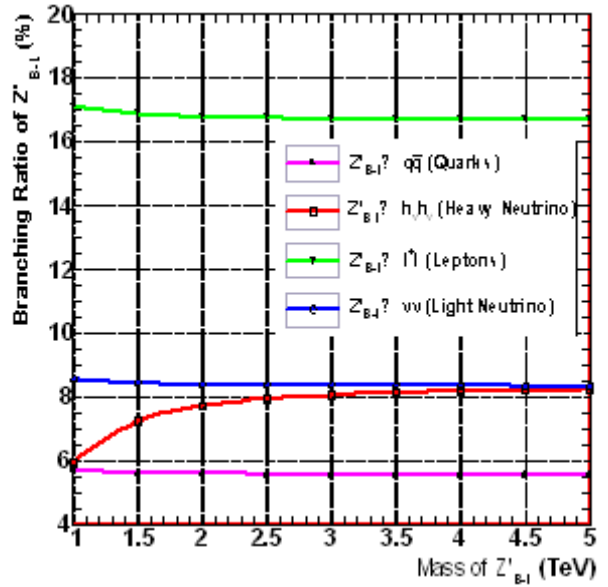

**FIG. 3.** Branching ratios for $Z'_{B-L}$ as a function of $M_{Z'_{B-L}}$ for $M_{h_v}$ = 200 GeV

### 3. Total width of $Z'_{B-L}$

$Z'_{B-L}$ boson decays only to fermions at tree-level and its width is given by the following expression

$$\Gamma(Z'_{B-L} \to f\bar{f}) = \frac{M_{Z'_{B-L}}}{12\pi} C_f (v^f)^2 [1+2\frac{m_f^2}{M_{Z'_{B-L}}^2}]\sqrt{1-\frac{4m_f^2}{M_{Z'_{B-L}}^2}} \qquad (4)$$

Where $m_f$ is the mass and $C_f$ the number of colors for the fermion type f. In figure 4 we present the total decay width of the $Z'_{B-L}$ as a function of $Z'_{B-L}$ mass for fixed values of g". Figure 5 presents the total decay width of the $Z'_{B-L}$ as a function of g" for fixed values of $Z'_{B-L}$ mass

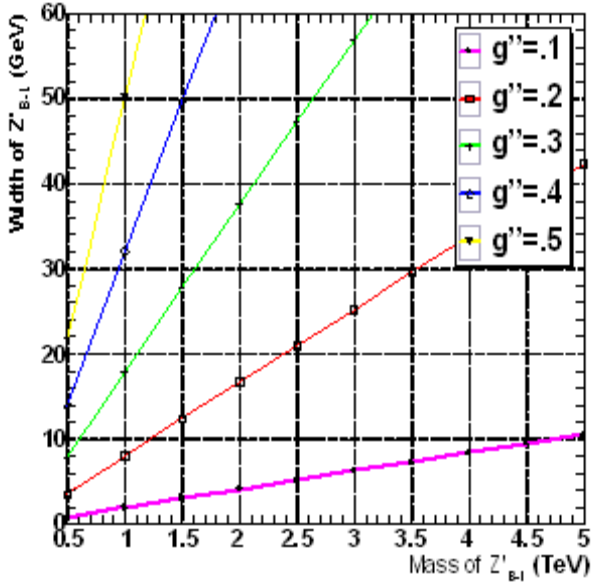

**FIG. 4.** Total width for $Z'_{B-L}$ as a function of mass $Z'_{B-L}$ for fixed values of g"

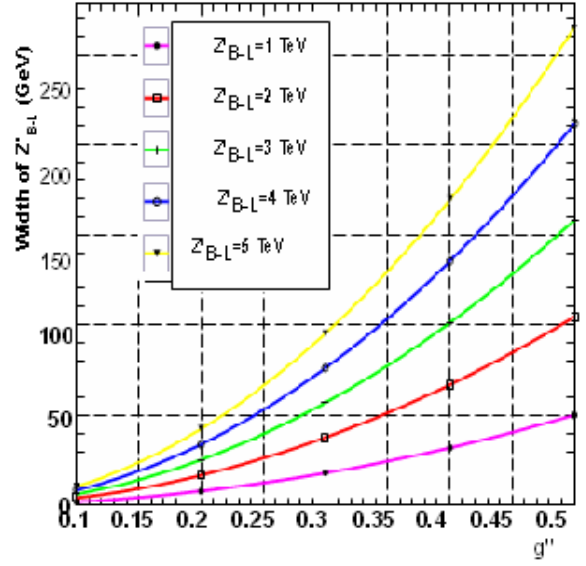

**FIG. 5.** Total width for $Z'_{B-L}$ as a function of g" for fixed values of $Z'_{B-L}$ mass

From figures 4 and 5 we note that the total width of a $Z'_{B-L}$ gauge boson varies from a few hundreds of GeV over a mass range of $0.5 < Z'_{B-L} < 5$ TeV depending on the value of g". The decay widths of $Z'_{B-L} \to ff$ in this model are then given by:

$$\Gamma(Z'_{B-L} \to l^+ l^-) \approx \frac{(g'' Y^l_{B-L})^2}{24\pi} m_{Z'_{B-L}} \quad (5)$$

$$\Gamma(Z'_{B-L} \to q\bar{q}) \approx \frac{(g'' Y^q_{B-L})^2}{8\pi} m_{Z'_{B-L}}$$
$$(1 + \frac{\alpha_s}{\pi}), q \equiv b, c, s \quad (6)$$

$$\Gamma(Z'_{B-L} \to t\bar{t}) \approx \frac{(g'' Y^t_{B-L})^2}{8\pi} m_{Z'_{B-L}}$$
$$(1 - \frac{4m_t^2}{m^2_{Z'_{B-L}}})^{\frac{1}{2}} (1 + \frac{\alpha_s}{\pi} + O(\frac{\alpha_s m_t^2}{m^2_{Z'_{B-L}}})) \quad (7)$$

The main switches for Initial State Radiation, Final Sate Radiation and multiple interactions are on. Figure 4 presents the total width for $Z'_{B-L}$ as a function of mass $Z'_{B-L}$ for fixed values of g" .Here, we used CM energy of LHC 14MeV. Figure 5 presents the total width for $Z'_{B-L}$ as a function of g" for fixed values of $Z'_{B-L}$ mass. We note that the total width of $Z'_{B-L}$ increase with increasing the mass of $Z'_{B-L}$.

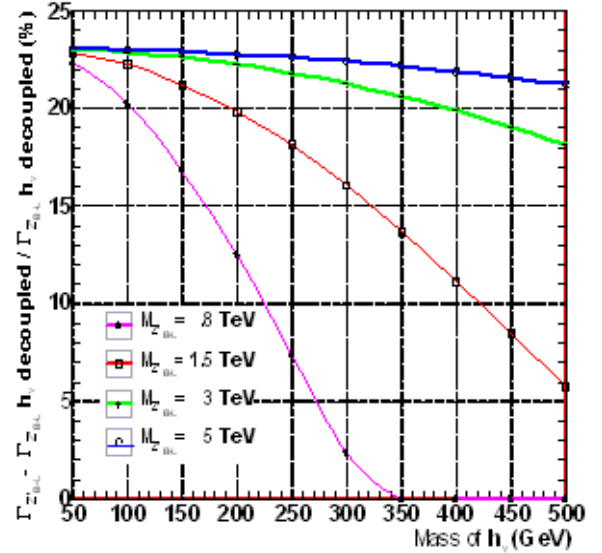

**FIG. 6.** Total width for $Z'_{B-L}$ as a function of heavy neutrino mass for fixed values of $M_{Z'_{B-L}}$ and g" = 0.3

In figure 6 we present the relative variation of the total width as a function of the heavy neutrino mass for different values of $M_{Z'_{B-L}}$ and for g"= .3. We note the importance of taking into consideration the heavy neutrino since their relative contribution to the total width can be as large as 20% where $\Gamma_{Z'_{B-L}}$ is the total width of $Z'_{B-L}$ which includes all decay channels (also decay channel of heavy neutrino ) whereas $\Gamma_{Z'_{B-L}}h_v$ decoupled includes all decay channels except decay channel of heavy neutrino is turn off.

## B. Detection of $Z'_{B-L}$ signal at LHC

### 1. Dielectron angular distribution

In addition to the dilepton invariant mass $M_{e^+e^-}$ analysis, it has been shown that the angular distribution of the dilepton events [10] can also be used to test the presence of a $Z'_{B-L}$ boson by detecting its interference with the SM Z boson. The massive resonance search technique ($M_{e^+e^-}$ analysis) is extended to include dilepton angular information to detect $M_{e^+e^-} \rightarrow e^+ + e^-$, so we will use the dielectron angular distribution $\cos(\theta^*)$ where $\theta^*$ is the angle in the dielectron rest reference frame between the negative electron and the incident incoming quark. PYTHIA8 gives $\theta$ only in Lab frame but we use $\theta^*$ here which is in rest frame so we must convert from lab frame to rest frame to get $\theta^*$ by using boost vector. We define two additional reference frames: (a) the colliding proton CM frame denoted by $O$ (this frame is identical to the laboratory frame) and (b) The rest frame of the dilepton system denoted by $O^*$. The dilepton system is boosted along the beam axis. The z-axis is chosen as the direction of one of the beams, and it is then identical for $O$ and $O^*$ frames. It should be noted that there is a sign ambiguity in the measurement of $\cos(\theta^*)$ since for a particular event, there is no information about whether the incoming quark comes from the positive or negative z directions [11]. Instead, it is useful to consider the quantity $\cos(\theta^*_\beta)$, where $\theta^*_\beta$ is the angle between the dilepton system boost $\vec{\beta}$ (relative to the $O$ frame) and the lepton direction

$$\cos(\theta^*_\beta) = \frac{\vec{P}^*_1 \cdot \vec{\beta}}{|\vec{P}^*_1| \cdot |\vec{\beta}|} \quad (8)$$

where the boost vector is

$$\vec{\beta} = \frac{\vec{P}_{l^+} + \vec{P}_{l^-}}{E_{l^+} + E_{l^-}} \quad (9)$$

In order to obtain $\vec{P}^*_1$ the boost vector of the dilepton system should be found and the transformation to the $O^*$ frame should be performed. Figure 7.gives the angular distribution of dielectron of $Z'_{B-L}$ boson decay as a function of $\cos(\theta^*)$.

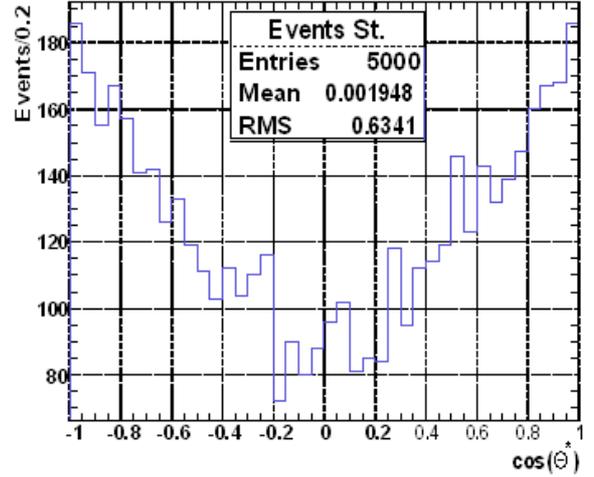

**FIG. 7.** Angular distribution of dielectron of $Z'_{B-L}$ boson decay where forward electrons have $\cos(\theta^*) > 0$ and backward electrons have $\cos(\theta^*) < 0$

### 2. Dielectron asymmetry $A_{FB}$

In the process $qq \rightarrow Z'_{B-L} \rightarrow ll$ where $Z'_{B-L}$ boson has both vector and axial vector couplings to the fermions. These couplings create an asymmetry in the momentum of the electron visible in the polar angle of the lepton pair's center of mass frame. This polar angle which is measured in the center of mass frame of the leptons is typically referred to as the Collins Soper frame. The angular cross section measured in this frame is given by:

$$\frac{d\sigma(q\bar{q} \rightarrow Z'_{B-L} \rightarrow l\bar{l})}{d\cos(\theta^*)} = A(1 - \cos^2(\theta^*)) + B\cos(\theta^*) \quad (10)$$

Here $\theta^*$ is the emission angle of the electron relative to the quark momentum in the lepton's center of mass frame(see figure 8). The constants A and B are determined by the weak isospin and charge of the incident quarks as well as the mass of the dilepton pair. From this cross section, it is convenient to define $N_f$ as the number of events whose $\theta^*$ is positive, and $N_b$ as the number of events whose $\theta^*$ is negative. The asymmetry can then be written as:

$$AFB = \frac{\left[\frac{d\sigma}{d\cos(\theta^*)}\right]_{>0} - \left[\frac{d\sigma}{d\cos(\theta^*)}\right]_{<0}}{\left[\frac{d\sigma}{d\cos(\theta^*)}\right]_{>0} + \left[\frac{d\sigma}{d\cos(\theta^*)}\right]_{<0}}$$

$$= \frac{3}{8} \times \frac{B}{A} = \frac{N_f - N_b}{N_f + N_b} \quad (11)$$

This is the general form of the asymmetry.

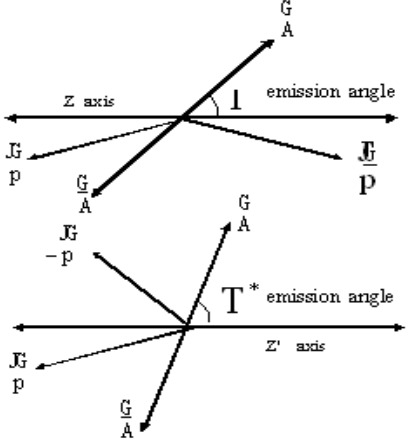

**FIG. 8.** The lepton's center of mass frame and the Collins-Soper frame.

Collins and Soper noted that an ambiguity exists in the determination of the emission angle $\theta^*$ when considering

$$q\bar{q} \to l\bar{l}$$

This Drell-Yan process is quite simple [12] when the incoming quarks have no transverse momentum. In such a case, the emission angle is determined by the angle the electron makes with the proton beam and thus the incoming quark momentum vector. However, since circular acceleration implies a certain amount of transverse momentum by construction, an ambiguity arises. Since the quark's individual momenta cannot be measured, the momenta boosted into the center of mass frame of the leptons are even more difficult to separate. Consequently, the dependence of the transverse momentum must be minimized. The polar $\theta^*$ axis is the bisector of the proton beam momentum and the negative (-) of the antiproton beam momentum when the two are boosted into the center of mass frame of the leptons. In so doing, the dependence on the transverse momentum of the incoming quark pair is minimized.

| $Z'_{B-L}$ mass (GeV) | Forward Electrons $N_F$ $\cos(\theta^*) > 0$ | Backward Electrons $N_B$ $\cos(\theta^*) < 0$ | Asymmetry $A_{FB}$ |
|---|---|---|---|
| 1500 | 1531 | 1469 | 0.020 |
| 1400 | 1520 | 1480 | 0.013 |
| 1300 | 1470 | 1530 | − 0.02 |
| 1200 | 1470 | 1530 | − 0.02 |
| 1100 | 1470 | 1530 | − 0.02 |
| 1000 | 1482 | 1518 | − 0.02 |
| 900 | 1531 | 1469 | 0.020 |
| 800 | 1498 | 1510 | − 0.004 |
| 700 | 1524 | 1476 | 0.016 |
| 600 | 1529 | 1471 | 0.019 |
| 500 | 1500 | 1500 | 0.000 |

**TABLE II.** The numbers of forward and backward electrons produced from $Z'_{B-L}$ decay in the range of $Z'_{B-L}$ mass 500 GeV to 1500 GeV and also asymmetry calculation.

Table II shows the numbers of forward and backward electrons produced from $Z'_{B-L}$ decay in the range of $Z'_{B-L}$ mass 500 GeV to 1500 GeV and also asymmetry calculation.

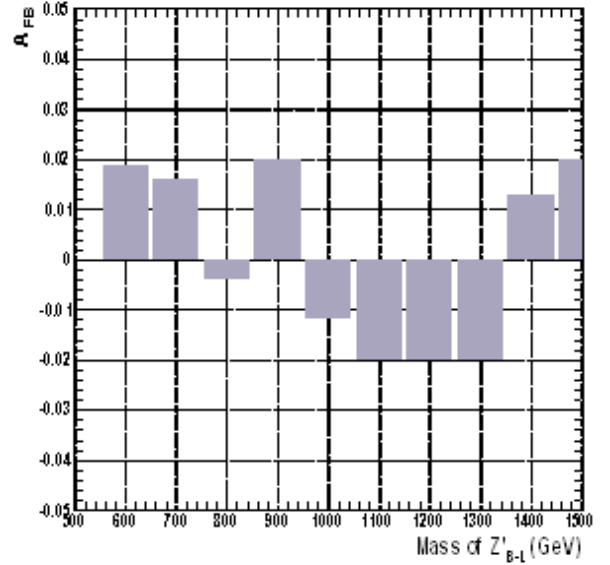

**FIG. 9.** Dielectron asymmetry distribution for forward and backward electrons for various values of $Z'_{B-L}$ masses.

Figure 9. shows the dielectron asymmetry distribution for forward and backward electrons for various values of $Z'_{B-L}$ masses.

*3. Drell Yan background events*

The histogram in figure 10 shows the generated dielctron events for a reconstructed mass of 50 to 700 GeV for SM $Z^o$. There is a peak centered on the 100 GeV for 6000 events were generated by PHYTHIA8. This peak, or resonant signal corresponds to the production of a $Z^o$ with a mass of 91.188 GeV. This process is called the Drell-Yan spectrum and dielectron are produced from $Z^o$ decay and they act as SM background for $Z'_{B-L} \to e^+ + e^-$ process

$$q\bar{q} \to Z^o \to e^+ + e^-$$

Here q is a quark from an incoming proton and it is annihilated with its antiparticle q from another incoming proton and produces a $Z^o$ which then decays into two dielec-

tron. The reconstructed mass of $Z^o$ will be calculated according to the equation:

$$M_z^2 = M_{e^+}^2 + M_{e^-}^2 + 2E_{e^+}E_{e^-} - 2P_{e^+}P_{e^-} \quad (12)$$

### 4. Dielectron invariant mass

Now, we will use the dilepton invariant mass $M_{e^+e^-}$ from the dilepton events [13] to test the presence of a $Z'_{B-L}$ boson at LHC through the massive resonance search technique ($M_{e^+e^-}$ analysis) beside the last method of dilepton angular information to detect $Z'_{B-L} \to e^+ + e^-$ which we have used before. By using PYTHIA8, ROOT and ROOFIT we created a series of $Z'_{B-L}$ signals with masses from 500 GeV/$c^2$ to 1500 GeV/$c^2$ as in table 3, then the mass of $Z'_{B-L}$ is reconstructed from the energy and momenta of the selected dielectron at coupling constant equal g"=0.1 [14]. We generated 5000 events for every signal mass where the backgrounds Drell-Yan events and desired signal events are selected by applying a number of selection cuts on all events samples(5000 events).

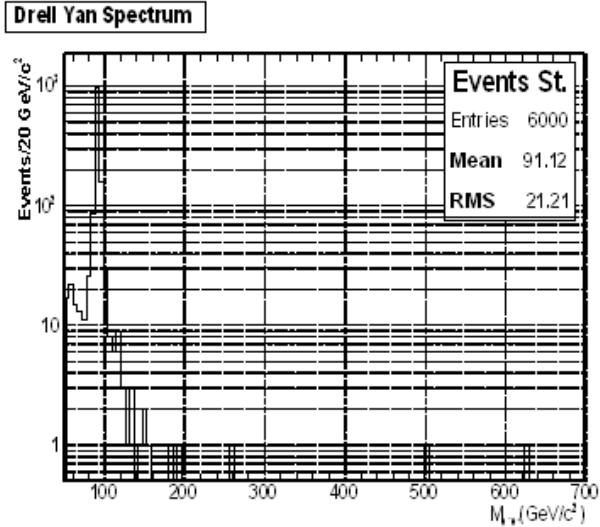

**FIG. 10.** The Drell-Yan background electrons produced from $Z^o$ boson for quark and antiquark annihilation.

1- Transverse energy of selected electrons $E_T > 100$ GeV.

2- The selected electrons must be in the central or in the forward regions of $|\eta| < 1.442$ or $1.566 < |\eta| < 2.5$ then we choose the two highest-energy electrons where $\eta$ is the pseudo rapidity of emitted electrons which describes the angle of a particle relative to beam axis.

$$\eta = -\ln\left[\frac{\theta}{2}\right] \quad (13)$$

$\theta$ is the angle between the particle momentum vector P and the beam axis(Fig.11).

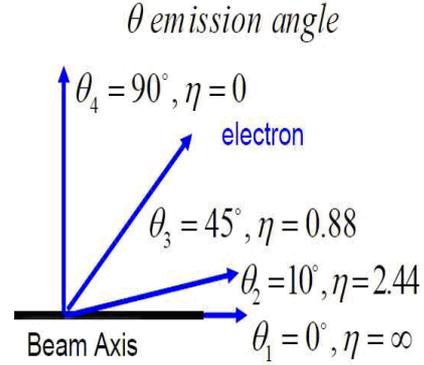

**FIG. 11.** The Pseudo rapidity of produced electrons of $Z'_{B-L}$ decay

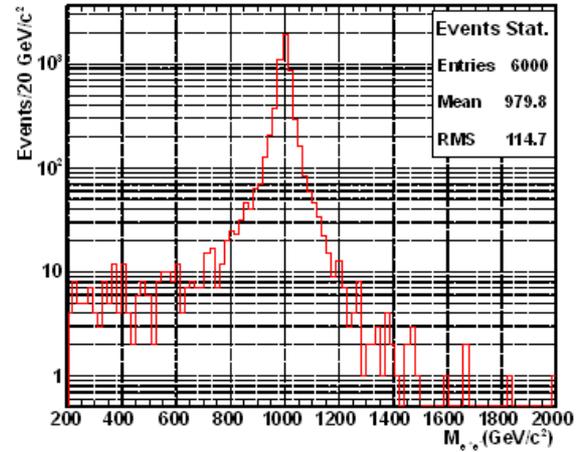

**FIG. 12.** The reconstruction for $Z'_{B-L}$ mass from produced dilepton invariant masses at $Z'_{B-L}$ mass 1 TeV.

From figure 12 we find a peak at 1 TeV which is the reconstruction mass of $Z'_{B-L}$ from the electron-positron invariant masses. This allowed us to compare the signal generated events with the Drell Yan background from figure 10.

| $Z'_{B-L}$ mass (GeV) | Expected events $N_{exp}$ | Mass window (GeV) | Signal events $N_S$ | Background events $N_B$ | Significance S |
|---|---|---|---|---|---|
| 1500 | 28.24 | 550 | 28.11 | 0.128 | 15.76 |
| 1400 | 38.36 | 520 | 38.25 | 0.102 | 19.40 |
| 1300 | 52.87 | 480 | 52.70 | 0.142 | 22.80 |
| 1200 | 73.30 | 460 | 73.10 | 0.206 | 26.70 |
| 1100 | 106.2 | 420 | 105.913 | 0.286 | 32.31 |
| 1000 | 157.2 | 450 | 156.7 | 0.424 | 39.30 |
| 900 | 237.9 | 500 | 237.2 | 0.639 | 48.38 |
| 800 | 375.5 | 480 | 374.4 | 1.013 | 60.70 |
| 700 | 641.8 | 500 | 640.07 | 1.720 | 79.40 |
| 600 | 1116 | 400 | 1113.01 | 2.980 | 104.8 |
| 500 | 2180 | 320 | 2174 | 5.8489 | 146.5 |

**TABLE III.** The expected number of events, signal events for $Z'_{B-L}$, background events, and Significance calculation at g"=0.1 using PYTHIA8.

### C. Properties of $Z'_{B-L}$

*1. Luminosity*

For each generated signal, PYTHIA8 calculated the cross section for each process. This is important because it allows us to scale our generated events to what an actual signal would look like given the luminosity [15]. The luminosity is the number of events collision per unit area in an accelerator. Therefore, the number of expected events $N_{exp}$ can be determined by the formula : $N_{exp} = L_\sigma$ where $N_{exp} = N_S + N_B$ and σ is the PYTHIA8 cross section of generated events and L is the luminosity and $N_S$ is the number of signal events and $N_B$ is the number of background events. Figure 13 give the required luminosity for $Z'_{B-L}$ observation at LHC for different masses at $Z'_{B-L}$. We note that the value of the luminosity increases by increasing $Z'_{B-L}$ mass.

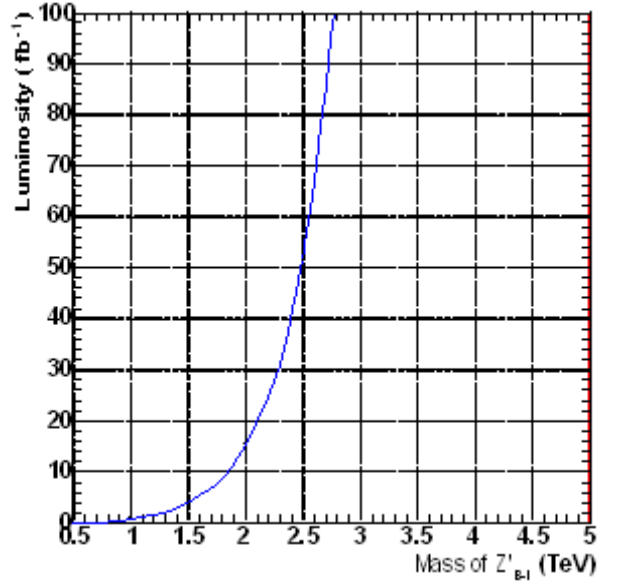

**FIG. 13.** Luminosity required for observation $Z'_{B-L}$ as a function of $Z'_{B-L}$ mass for g"=0.2 at the CM energy of LHC √s = 14 TeV

*2. Significance calculations*

To calculate the significance, each reconstructed mass is fitted by a Gaussian using the ROOFIT package and using the standard deviation with 3σ mass window around the fitted peak for example from figure 12 at 1000 GeV peak the mass windows is 450 GeV then σ = 75 and μ = 1000. We will integrate the Gaussian from 550 to 1450 to get the fraction of signal event, then we will multiply this fraction by the total expected events $N_{exp}$ to get the number of signal event $N_S$ then we can calculate the background events $N_B$ by:

$$N_B = N_{exp} - N_S \quad (14)$$

The significance formula is:

$$S = \sqrt{2\left[(N_S + N_B)\ln(1 + \frac{N_S}{N_B}) - N_S\right]} \quad (15)$$

Where $N_S$ is the number of signal events and $N_B$ is the number of background events. Table III shows a summary of signal events and background events for different values of $Z'_{B-L}$ mass and the corresponding significance. Figure 14 shows the signal significance as a function of the mass of $Z'_{B-L}$ for g" = 0.2.

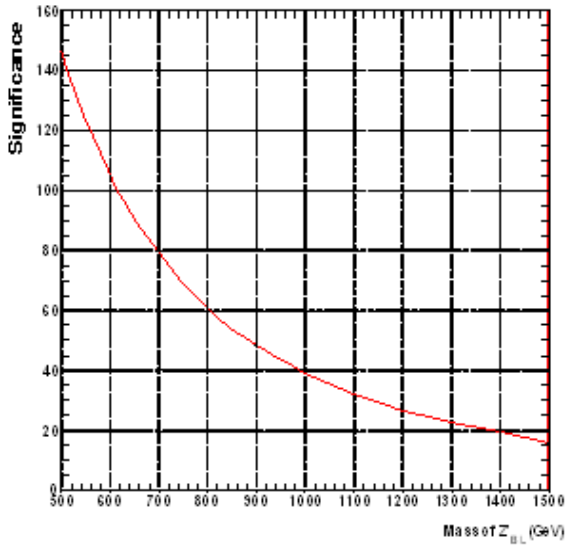

**FIG. 14.** Signal significance as a function of the mass for $Z'_{B-L}$ for g" = 0.2

*3. Transverse momentum*

Figure 15 show the transverse momentum $P_T$ distributions of positron and electron pairs in invariant mass region $500 < M_{e^+e^-} < 1500$ GeV produced from $Z'_{B-L}$ decay. The results are obtained using PYTHIA8 of proton-proton collisions at a center-of-mass energy of 14 TeV at LHC and g"=0.2. The distributions are measured over the ranges |η| < 1.442 or 1.566 < |η| < 2.5 and the transverse energy of selected electrons is $E_T > 100$ GeV. The distributions for y and $P_T$ are normalized by total cross sections within acceptance regions described are shown in figures 16 and 17.

Figures 18 and 19 show the measurement of the rapidity y where the transverse-momentum $P_T$ distributions and rapidity of the $Z'_{B-L}$ boson provide a new information about the dynamics of proton collisions at high energies and the $P_T$ spectrum provides a better understanding of the underlying collision process at low transverse momentum.

The measurements of the rapidity and transverse momentum spectra are based on samples over boson events reconstructed in each dilepton decay mode and collected using high $P_T$ single lepton. For the $Z'_{B-L}$ boson candidates selected pairs of leptons are required to have a reconstructed invariant mass in the range $500 < M_{e^+e^-} < 1500$ GeV. The two electrons candidates with the highest $P_T$ in the event are used to reconstruct a $Z'_{B-L}$ candidate. Electrons are required to have $E_T >100$ GeV and |η| < 1.442 or 1.566 < |η| < 2.5

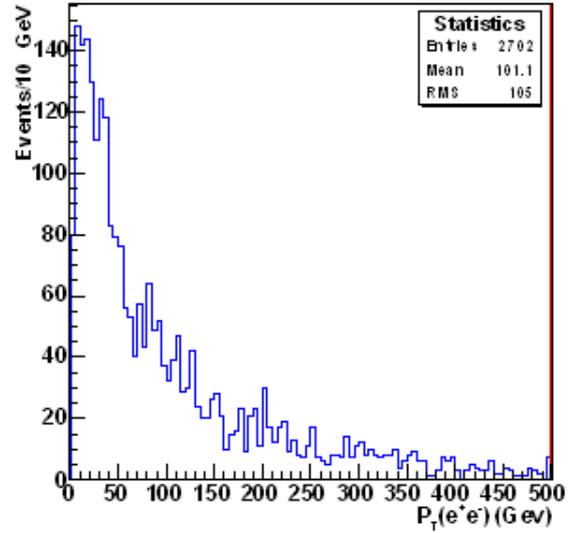

**FIG. 15.** Transverse momentum for dielectron for mass of $Z'_{B-L}$ = 1 TeV and energy of LHC = 14 TeV and g" = 0.2

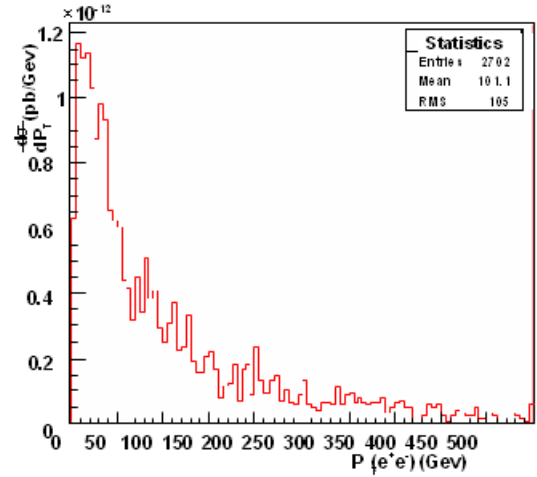

**FIG. 16.** Differential cross section as a function of dielctron transverse momentum at g" = 0.2 and energy of LHC = 14 TeV

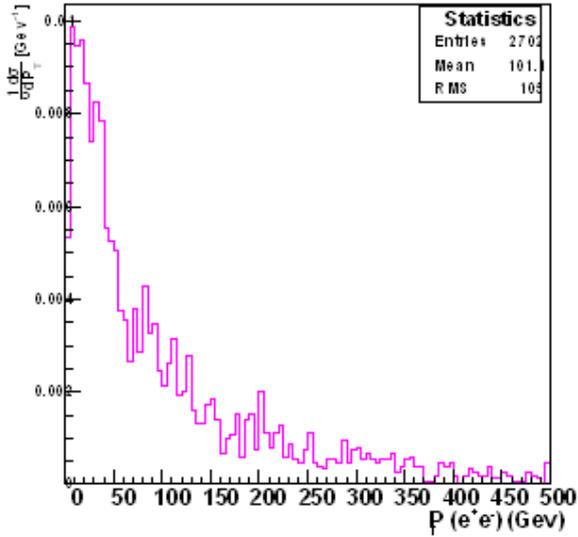

**FIG. 17.** The normalized differential cross section for $Z'_{B-L}$ boson as a function of transverse-momentum of dielectron

**4. Rapidity of $Z'_{B.L}$**

The rapidity is defined as

$$y = \frac{1}{2}\ln\frac{(E+P_T)}{(E-P_T)} \quad (16)$$

where E is the energy of the $Z'_{B-L}$ candidate and $P_T$ is its longitudinal momentum along the anticlockwise beam axis (the z axis of the detector). The y and $P_T$ of $Z'_{B-L}$ are determined from the leptons momenta. The measured differential dielectron cross sections are normalized to the inclusive Z cross section. The differential cross section is determined in each y or bin by subtracting from the number of detected events in a bin the estimated number of background events. The distributions are corrected for signal acceptance and efficiency and for the effects of detector resolution and electromagnetic final-state radiation (FSR) using an unfolding technique based on the inversion of a response matrix. The final result takes into account the bin width and is normalized by the measured total cross section.

The distribution of $Z'_{B-L}$ bosons is symmetric about y = 0 and therefore the appropriate measurement is the distribution as a function of the absolute value of rapidity The measurement is normalized to the total cross section

$$\frac{1}{\sigma}\frac{d\sigma}{d|y|} \quad (17)$$

and σ is the total cross section is determined by 2702 events

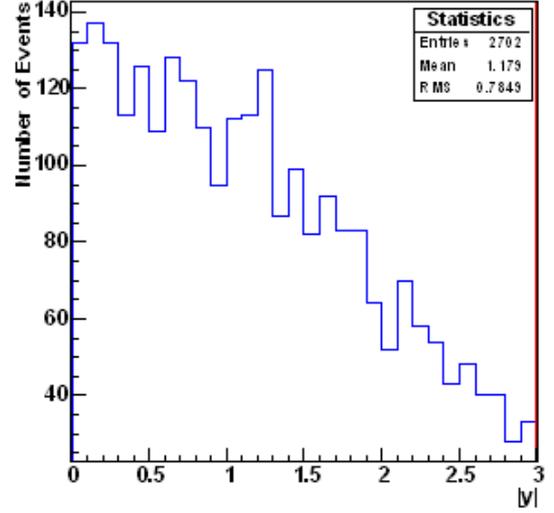

**FIG. 18.** Absolute value of rapidity for dielectron for mass of $Z'_{B-L}$ = 1 TeV and energy of LHC=14 TeV and g''= 0.2

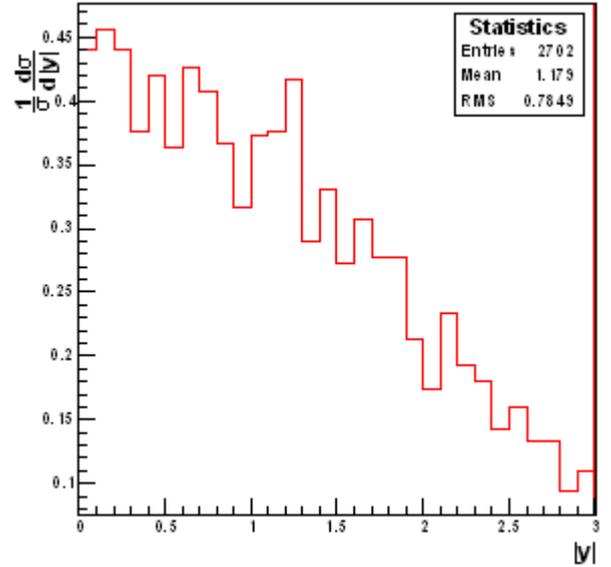

**FIG. 19.** The normalized differential cross section for $Z'_{B-L}$ as a function of the absolute value of rapidity of dielectron for mass of $Z'_{B-L}$ = 1 TeV at CM energy of LHC = 14 TeV and g'' = 0.2

### III. CONCLUSIONS

In this work we have presented the LHC potential to discover a heavy neutral massive gauge boson $Z'_{B-L}$ in B-L extension of SM model using MC programs where we have simulated the production of $Z'_{B-L}$ for different center of mass energies at LHC for various values of coupling constant g" and also presented all possible branching ratios of $Z'_{B-L}$ to different decay channels to fermions and presented the total width of $Z'_{B-L}$. We have used both dielectron angular distribution and dielectron invariant mass to detect $Z'_{B-L}$ signal at LHC. Finally we calculated the luminosity, significance, dielectron transverse momentum, rapidity and differential cross section. All these signatures predict the existence of a new gauge boson $Z'_{B-L}$ at LHC in the mass range 1 TeV to 1.5 TeV.

## ACKNOWLEDGEMENTS

It is a pleasure to thank T. Sjostrand for useful discussions of PYTHIA, L. Basso and C. Duhr for useful discussions of B-L model and J. Alwall for useful discussions of Mad-Graph5/MadEvent.